\documentclass[10pt,twocolumn,letterpaper]{article}
\usepackage{booktabs}
\usepackage{makecell}
\usepackage{wacv}              

\definecolor{wacvblue}{rgb}{0.21,0.49,0.74}
\usepackage[pagebackref,breaklinks,colorlinks,allcolors=wacvblue]{hyperref}
\usepackage{booktabs}

\def\wacvPaperID{} 
\def\confName{WACV}
\def\confYear{2026}

\usepackage[table]{xcolor}
\usepackage{adjustbox}

\title{
EZBlender: Efficient 3D Editing with Plan-and-ReAct Agent 
 
}

\author{Hao Wang$^1$\thanks{Corresponding author}, Wenhui Zhu$^3$, Shao Tang$^2$, Zhipeng Wang$^2$, Xuanzhao Dong$^3$, Xin Li$^3$,\\ Xiwen Chen$^1$, Ashish Bastola$^1$, Xinhao Huang$^4$, Yalin Wang$^3$, and Abolfazl Razi$^1$\\
$^1$Clemson University, SC, USA\\
$^2$LinkedIn Corporation, CA, USA \\
$^3$Arizona State University, AZ, USA\\
$^4$Pace University, NY, USA\\
{\tt\small \{hao9, xiwenc, abastol, arazi\}@clemson.edu, \{shatang, zhipwang\}@linkedin.com,}\\
{\tt\small \{wzhu59, xdong64, xinli38, ylwang\}@asu.edu, xh87220n@pace.edu}
}

\begin{document}
\maketitle

\begin{abstract}
As a cornerstone of the modern digital economy, 3D modeling and rendering demand substantial resources and manual effort when scene editing is performed in the traditional manner. 
Despite recent progress in VLM-based agents for 3D editing, the fundamental trade-off between editing precision and agent responsiveness remains unresolved.
To overcome these limitations, we present EZBlender, a Blender agent with a hybrid framework that combines planning-based task decomposition and reactive local autonomy for efficient human–AI collaboration and semantically faithful 3D editing. 
Specifically, this unexplored Plan-and-ReAct design not only preserves editing quality but also significantly reduces latency and computational cost.
To further validate the efficiency and effectiveness of the proposed edge-autonomy architecture, we construct a dedicated multi-tasking benchmark that has not been systematically investigated in prior research. In addition, we provide a comprehensive analysis of language model preference, system responsiveness, and economic efficiency. 
Our example dataset and code are available at:
\url{https://github.com/Aztech-Lab/EZ_Blender}
\vspace{-0.5cm}
\end{abstract}


\section{Introduction}
\label{sec:intro}

Three-dimensional (3D) modeling has become a cornerstone of contemporary digital entertainment and creative industries at large \cite{zhuang2023dreameditor,karim2024free,SODA}. Virtually every major domain of visual media depends on the expressive power of 3D modeling \cite{bhuyan2024gamestreamsr,banfi2024gaming}. 
However, beneath its artistic appeal lies a reality of formidable industrial complexity \cite{li2024advances}. 3D modeling is rarely the work of a single operation; it is a sprawling ecosystem of interconnected pipelines that span design, geometric modeling, material authoring, animation, lighting, physics simulation, rendering, and scene composition \cite{flavell2011beginning,Latent3DGS,BezierSplat}. Each of these stages demands dedicated expertise, specialized software, and extensive coordination \cite{hosen2019mastering}. 
This comprehensive workflow creates a steep barrier to entry, 
where small teams and independent creators, despite having creative vision, often lack the infrastructure to execute it at scale \cite{AHMAD20171,cai20233description}. 

\begin{figure}
    \centering
    \includegraphics[width=0.9\linewidth]{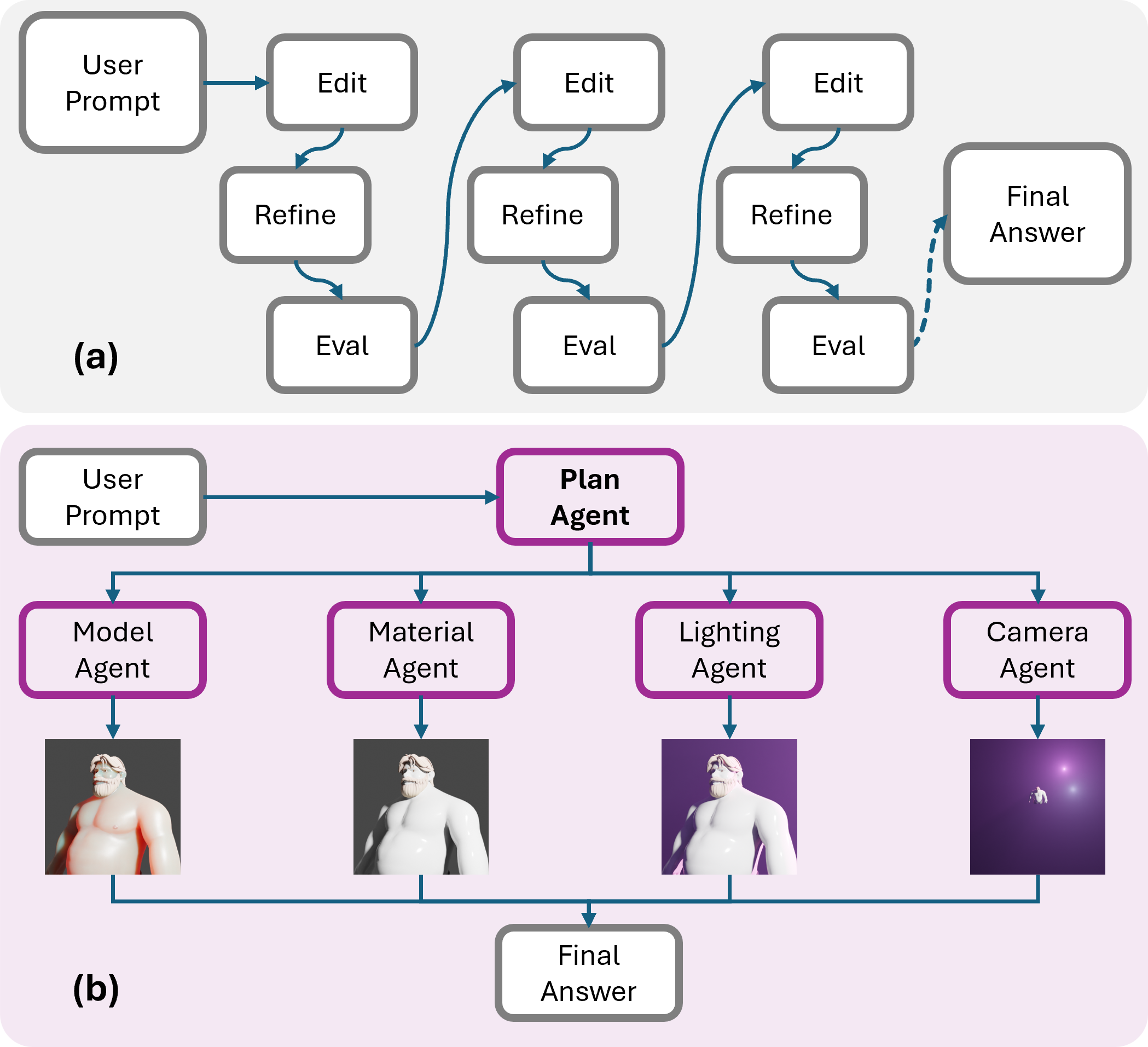}
    \caption{The design concept of 3D editing agents. (a) The leap-tune design implemented in BlenderAlchemy\cite{huang2024blenderalchemy}; (b) The proposed Plan-and-ReAct agent design.}
    \vspace{-0.5cm}
    \label{fig:cover}
\end{figure}

Beyond this structural bottleneck lies a deeper and more intrinsic challenge: the arduous process of turning abstract concepts into production-ready assets \cite{raistrick2023infinite,liu2025worldcraft}. Unlike textual or 2D creative workflows \cite{SODA,FlowStraighter}, where ideation and expression can be relatively immediate, 3D modeling requires a gradual process of externalization \cite{hamdani20233d,nie2025ermv}. 
This gap between conceptual imagination and executable representation represents a fundamental barrier in 3D creation:
the translation of human intent into computationally actionable form.

Nevertheless, the recent advent of large language models (LLMs) has begun to disrupt this long-standing stalemate \cite{NanoBananaAI}. With their unprecedented contextual reasoning ability, LLMs can decompose inherently complex tasks into clear, manageable sub-problems, thereby alleviating the cognitive burden that traditionally accompanied large-scale 3D workflows \cite{michel2023object,sun20253d,nie2025ermv}. 
This not only lowers the technical barrier of entry but also accelerates workflows by reducing the workload from repetitive tasks, creating the possibility of exponential efficiency gains across the creative industries.
More recently, the rapid advances of vision–language models (VLMs) have fundamentally altered the landscape of creative workflows \cite{,jun2023shap,lin2023magic3d,lv2024gpt4motion}. 
In particular, natural language interaction is starting to serve as a critical bridge between traditional flat design workflows and emerging multi-dimensional modalities  such as VR, spatial stylus input, and haptic interfaces, which have massively improved efficiency in human–AI collaboration\cite{schmidgall2025agent,gollnersony,chen2025symbolic}.
Systems such as Manus\cite{Manus}, ChatGPT Agents\cite{OpenAIChatGPTAgent}, AutoGen\cite{wu2024autogen}, and CAMEL\cite{li2023camel} go well beyond single-task execution, demonstrating that modern agents are evolving into flexible ecosystems rather than fixed patterns, capable of coordinating multiple roles and handling intricate workflows.

BlenderGPT was among the first to bring LLMs into 3D editing, enabling users to control Blender directly through natural language \cite{BlenderGPT}. More recently, agentic designs such as BlenderAlchemy have demonstrated that coupling LLMs and VLMs with iterative feedback can achieve partially automated and more precise scene editing \cite{huang2024blenderalchemy}. These advances clearly showcase the potential of language- and vision-driven agents in graphics. 
However, despite the impressive capabilities of recent vision–language models, most existing workflows have not adequately addressed the fundamental trade-off between efficiency and performance. For instance, 
systems that optimize purely for efficiency often miss user intent in subtle ways, while those that focus heavily on refinement and accuracy incur high latency and break the interactive flow\cite{huang2024blenderalchemy}. 

To address this challenge, we introduce \textit{EZBlender}, a Blender \cite{blender} agent for natural language–driven 3D editing tasks. Inspired by modern film production pipelines \cite{alma9976203403606533}, EZBlender integrates principles from both ReAct \cite{yao2023react} and Plan-and-Execute \cite{PlanAndSolve} agent designs, and restructures them into a parallelized workflow, as shown in Figure \ref{fig:cover}. 

Specifically, a Planner Agent performs high-level reasoning to decompose the user’s intent into domain-aware semantic instructions, such as geometry updates or lighting shifts. 
Each instruction is then routed to a dedicated Sub-Agent. Based on its specific system prompt and local context, each sub-agent then acts as a domain expert, autonomously grounding coarse semantic instructions into executable hard constraints (e.g., specific shape key values) and generating the corresponding Blender code snippets. 
This unexplored Plan-and-ReAct design relinquishes autonomy to the endpoints, eliminating deep dependency chains and reducing redundant VLM calls, thereby avoiding the bottleneck of single-sequence threading. 
As demonstrated in our experiments, this new agent design also improves responsiveness by up to about 7 times faster than the previous agent layout \cite{huang2024blenderalchemy}, while preserving semantic fidelity to the user’s instructions, thereby effectively narrowing the gap between efficiency and performance.

\begin{figure*}[htbp]
    \centering
    \includegraphics[width=1\linewidth]{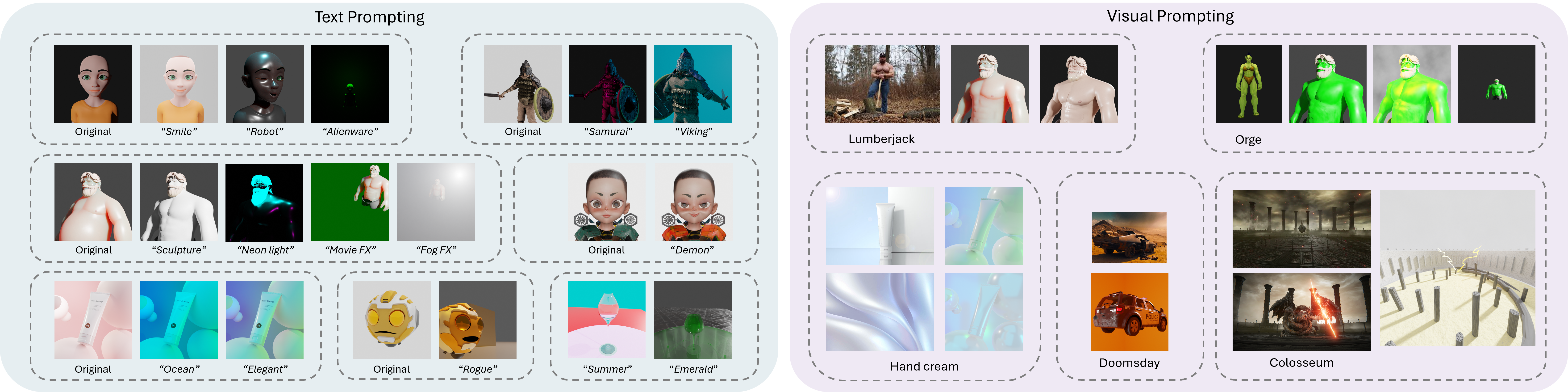}
    \caption{\textbf{Visualization results of our method.} 
    \textbf{(Left.)} Text-prompt-driven editing, where our agent applies textual instructions to manipulate 3D-rendered avatars and objects, including \textbf{attribute modifications} (e.g., skin color, accessories, identity), \textbf{stylistic transformations} (e.g., cartoon, pixel-art), and \textbf{scene/background changes}. 
    \textbf{(Right.)} Visual-prompt-driven editing, where reference images guide transformations, enabling \textbf{shape transfer}, \textbf{illumination adjustment}, \textbf{structural modifications}, and \textbf{scene-level re-rendering} (e.g., architectural edits, object replacement). 
}
    \label{fig:visual}
\end{figure*}

The main contributions of this work are straightforward:
\begin{itemize}
\item To improve overall system responsiveness, we propose a hybrid agent architecture that is specifically designed for the Blender application\cite{blender}. Our system parallelizes sub-task execution priorities, thereby reducing redundant inter-agent communication and mitigating bottlenecks introduced by individual components. 
    \item To improve the performance and success rate of task completion, we delegate autonomy to specialized sub-agents \cite{adk_python}. This design also enables more efficient atomic modifications, encourages the generation of scene variants, and significantly enhances the textual and visual alignment with the user prompt, as shown in Figure \ref{fig:visual}.
\item To systematically evaluate the precision and efficiency of AI agents in 3D editing tasks, we introduce a quantifiable multi-task benchmark for measuring task completion rate. The benchmark covers 85 episodes and five evaluation dimensions, each assessed using a contrastive learning model (CLIP \cite{radford2021learning}) to eliminate subjective bias. 

\end{itemize}



\section{Related Works}
\label{sec:related}

\begin{figure}
    \centering
    \includegraphics[width=1\linewidth]{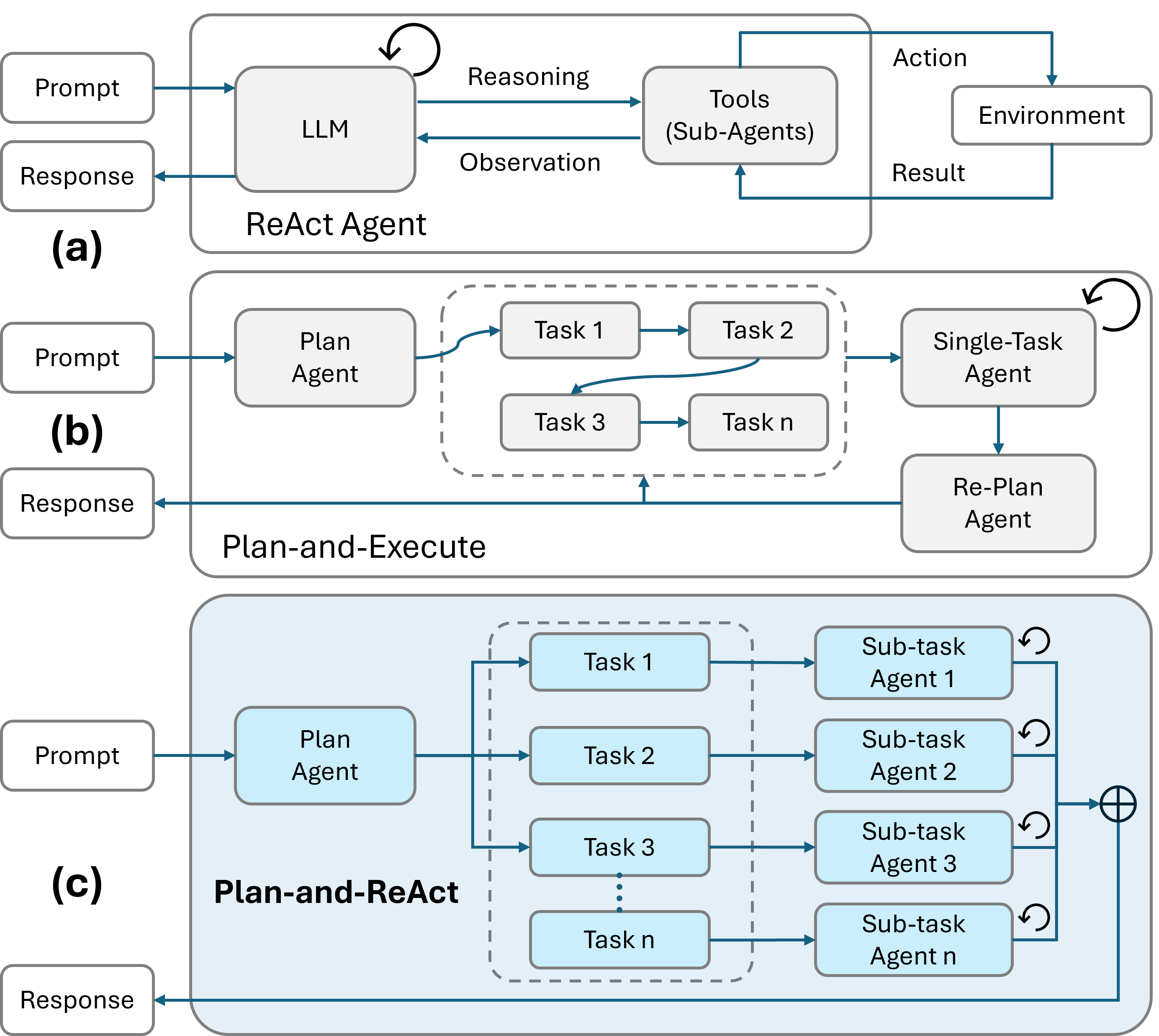}
    \caption{Agent design comparison. (a) The mainstream ReAct \cite{yao2023react} agent design; (b) The mainstream Plan-and-Execute agent design; (c) The proposed Plan-and-ReAct agent design optimized for 3D editing work.}
    \label{fig:react}
    \vspace{-0.5cm}
\end{figure}

\subsection{Prompt Engineering \& AI Agents}


Large language models (LLMs) and, more recently, vision–language models (VLMs) have enabled natural-language interfaces for a broad range of tasks \cite{wang2023voyager,liu2024deepseek}. Prompt engineering has emerged as a central paradigm for adapting LLMs to specialized domains without additional training \cite{wang2023voyager,schmidgall2025agent}. 
Specifically, early work improved downstream performance via template-based prompts that steer model behavior for question answering, captioning, and even code generation \cite{RetinalGPT,LLaDAMedV,chen2025symbolic,dong2025talk}. 
As model capabilities and modalities expanded, the community shifted from static prompts to agentic workflows, where LLMs/VLMs are orchestrated as autonomous agents capable of invoking tools and managing memory. 
For instance, systems such as AutoGPT \cite{AutoGPT} and Manus \cite{Manus} introduced pipelines that dynamically plan and iteratively execute toward a goal \cite{superdesign}.


As contemporary agent designs have become far more complex and diverse, the boundary between general and specialized agents is rapidly blurring, with numerous variants (e.g., Reflexion \cite{shinn2023reflexion}, Supervisor \cite{langgraphSupervisor}) extending the design space, and multi-agent collaboration further complicating the landscape. 
The debate between specialized agents and more generalized models is still ongoing, with no definitive conclusion yet.
However, despite this flourishing ecosystem, no agent framework has yet demonstrated a strong impact in the domain of 3D editing. A practical, domain-adapted 3D editing agent remains highly sought after.

\begin{figure*}[htbp]
    \centering
    \includegraphics[width=0.9\linewidth]{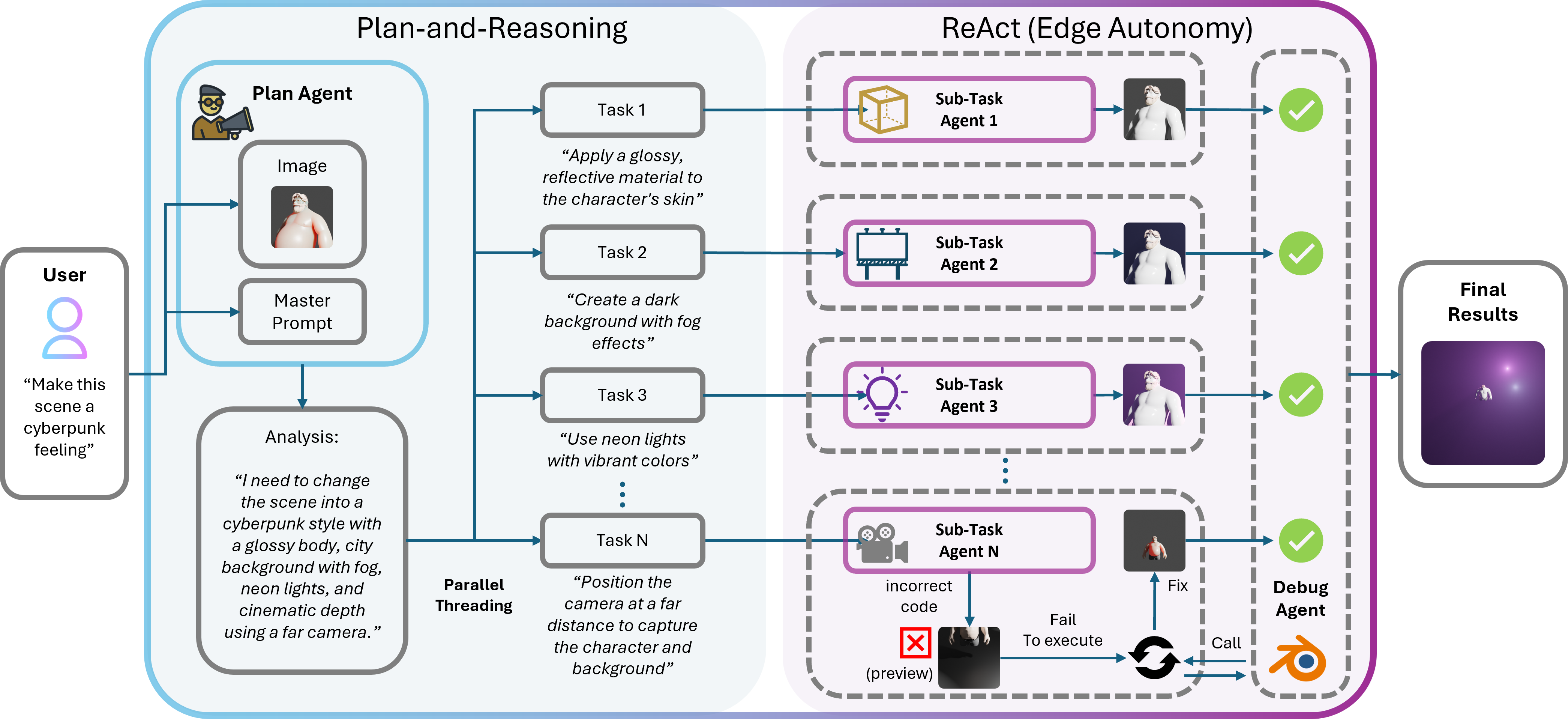}
    \caption{Proposed EZBlender framework. The planner agent takes user input and decomposes it into several sub-tasks with templated reasoning. Specialized sub-agents then generate the corresponding scene components and refine them through edge autonomy. The system outputs the final response once all components are completed.}
    \label{fig:ez_frame}
    \vspace{-0.25cm}
\end{figure*}


\subsection{LLM in 3D Editing}

Recent 3D generation methods that synthesize novel 3D assets primarily emphasize geometric and mathematical modeling capabilities \cite{gao2022get3d,lu2025ll3m,Tripo3D,he2024freestyle,yi2024gaussiandreamer}. 
In contrast, 3D editing pipelines focus on task-specific modifications via interactions with the environment \cite{BlenderGPT,lu2025ll3m}. These editing-oriented methods place greater emphasis on scene understanding, state-dependent manipulation, and executable control over pre-existing content \cite{huang2024blenderalchemy}.

In 3D applications such as Blender \cite{blender}, LLMs began to assume the roles of planning and tool-use  with the shift toward agentic design \cite{BlenderGPT}, as 3D editing values interpretability and editability \cite{lu2024advances}.
For instance, SceneCraft formalized text-to-scene generation by first planning a scene graph and then emitting Blender-executable code to instantiate objects and spatial relations, enabling layout-aware room-scale synthesis \cite{yang2024scenecraft}. 
In parallel, agent systems developed for game and interactive environments (e.g., level generation and embodied agents) further underscore how LLM planning and tool calls can support complex content-creation workflows \cite{wang2023voyager}.

\section{Methods}
\label{sec:method}

\subsection{Plan-and-Reasoning}

While the ReAct paradigm emphasizes a powerful general-purpose agent and Plan-and-Execute prioritizes task specialization, they both exhibit clear limitations: ReAct relies heavily on sequential reasoning, which leads to long dependency chains and error accumulation \cite{yao2023react}, while Plan-and-Execute suffers from high end-to-end latency and costly re-planning \cite{PlanAndSolve}. 

To improve the agent's responsiveness in 3D editing tasks, we adopt a hierarchical agent architecture for both semantic interpretation and task decomposition. 

\noindent \textbf{Semantic Disentanglement.}
User prompts in 3D editing are typically holistic and underspecified, often containing stylistic or high-level descriptions (e.g., ``\textit{Make the scene look cyberpunk}'') without explicit technical parameters. 
Thus, the Planner Agent $\mathcal{A}_{plan}$ must first interpret the user's intent:
\begin{align}
  \mathbf{U} = (T, I),
\end{align}
where $T$ is a natural-language prompt and $I$ is an optional visual reference \cite{radford2021learning}.
The goal of this stage is to extract domain-relevant semantic meanings before any task formulation.
We define the set of 3D editing domains as: $\mathcal{K} = \{ \text{geo}, \text{mat}, \text{light}, \text{cam}, \text{bg}\}$, then semantic disentanglement maps the user intent $U$ into domain-specific semantic factors:
\begin{align}
    \mathcal{Z} = \{ z_k \mid k \in \mathcal{K} \}, \quad 
z_k = \Phi_k(U),
\end{align}
where $\Phi_k$ is the semantic projection for domain $k$.
Each $z_k$ captures the interpreted meaning of the prompt for that domain, but is not yet an actionable instruction.
For instance, the abstract concept ``cyberpunk'' can be disentangled as:
\begin{align*}
z_{\text{light}} &= \text{``neon, high-contrast lighting''}, \\
z_{\text{mat}} &= \text{``wet, reflective metallic materials''}.
\end{align*}
This step preserves the artistic semantics of the user intent while avoiding prematurely binding to Blender-specific syntax.

\noindent \textbf{Task Decomposition.}
After semantic disentanglement, the planner converts $\mathcal{Z}$ into a structured set of domain-aware editing directives suitable for downstream sub-agents.
We define task decomposition as:
\begin{align}
\mathcal{S} = \mathcal{A}_{plan}(U) = \Psi(\mathcal{Z})
= \{ \delta_k \mid k \in \mathcal{K} \},
\end{align}
where each $\delta_k$ is a domain-specific editing directive.

Specifically, $\delta_k$ is expressed as semantic specifications rather than executable code to satisfy the following objectives:
(i) decouple high-level reasoning from Blender API implementation,
(ii) remain stable across Blender versions and add-on configurations,
and (iii) serve as clean, interpretable interfaces for various sub-agents.

This design preserves ReAct’s reasoning capacity to regularize ambiguous instructions into a compact specification and adopts Plan-and-Execute’s explicit task decomposition, but replaces a generic executor with domain-specialized sub-agents, as shown in Figure \ref{fig:ez_frame}.



\subsection{Edge Autonomy}

While Plan-and-Reasoning covers the planning side, the ReAct element lies at the endpoints: we decentralize evaluation and debugging from the planner to domain-specific sub-agents at the point of action. For each assigned fragment, the sub-agent runs a bounded propose–verify–refine cycle, as illustrated in Figure \ref{fig:ez_frame}.



Specifically, to ground the Planner-issued directive~$\delta_k$, each sub-agent generates its action through an LLM-based function:
\begin{align}
c_k \sim g_k(\delta_k; \tau_k),
\end{align}
where $\tau_k$ is the temperature parameter of LLMs which regulates the degree of freedom permitted during code generation.
However, before sampling $c_k$, the sub-agent must convert the semantic directive into a 
deterministic hard-constraint set:
\begin{align}
\mathcal{C}_k = h_k(\delta_k),
\end{align}
where $h_k$ extracts all non-negotiable requirements implied by $\delta_k$. For example:
\begin{align}
    \text{``blue lighting''} \rightarrow
&\big\{
\text{light.color} = (0.05, 0.25, 1.0)
\big\}, \\
\text{``red fog''} \rightarrow
&\big\{
\text{volume.color} = (1.0, 0.1, 0.1)
\big\}.
\end{align}
These constraints shape the final code, while the LLM-based sampling with temperature~$\tau_k$ determines how the constraint is realized.
Higher temperatures allow the sub-agent to deviate more freely from the surface form of the directive. For example, choosing a more complicated lighting option instead of a conservative preset, and vice versa.
This controllable edge autonomy design allows partial decision-making to be delegated to the execution end and encourages the generation of scene variants.

\noindent \textbf{Universal Debug Agent}
Nevertheless, with such autonomy, some generated scripts may contain execution errors (e.g., invalid node names, missing attributes, out-of-range parameters).  
To maintain robustness, each sub-agent may invoke a universal Debug Agent $\mathcal{A}_{dbg}$ whenever validation fails.
The correction pipeline follows:
\begin{align}
c^{(t+1)}_k = 
\begin{cases}
c^{(t)}_k, & \mathcal{V}(c^{(t)}_k) = \text{pass}, \\[4pt]
\mathcal{A}_{dbg}\!\left(c^{(t)}_k, \mathcal{V}(c^{(t)}_k)\right), & \text{otherwise},
\end{cases}
\end{align}
where $\mathcal{V}$ is our lightweight Blender-side validator, as shown in Figure \ref{fig:ez_frame}.
The Debug Agent applies generic, domain-agnostic repair strategies such as replacing unsafe parameter values with defaults, correcting missing Blender identifiers, and resolving reference errors (e.g., absent objects or collections).
Importantly, the Debug Agent is accessible to all sub-agents and does not require global re-planning.  

This bounded autonomy allows sub-agents to resolve small inconsistencies in situ, prevents upstream errors from propagating, and improves the single-pass task-completion rate without compromising scene coherence.


\section{Experiments \& Results}
\label{sec:result}
\subsection{Experimental Setup}

Since this work primarily focuses on 3D editing tasks, we evaluate the proposed EZBlender against existing Blender-based editing systems: BlenderGPT \cite{BlenderGPT} and BlenderAlchemy \cite{huang2024blenderalchemy}, for a fair comparison. Particularly, BlenderAlchemy is configured with the default 3×4 tree-search profile provided in the BlenderGym dataset.
For rigor, we adopt BlenderGym’s \cite{gu2025blendergym} 245 preset Blender scenes and follow its recommended rendering configuration: Cycles rendering engine (with CUDA support), 512×512 resolution, and two views per render cycle. Meanwhile, all Blender scenes follow BlenderGym’s default $\textbf{.blend}$ configurations with no post-hoc modifications or manual tuning.
All experiments are conducted under identical hardware: RTX 3090 (24 GB), 96 GB RAM, with an Intel Core i9-13900K CPU on a Windows 11 desktop.
In addition, all agents use GPT-4o as the LLM backend by default, unless otherwise specified.

Specifically, we evaluate three settings: 
(a) prompt-editing (text and visual), (b) multi-task editing, and (c) latency. User-side prompts are identical across methods, and all trials follow the same protocol with fixed task lists to ensure validity.

\begin{figure*}[htbp]
    \centering
    \includegraphics[width=1\linewidth]{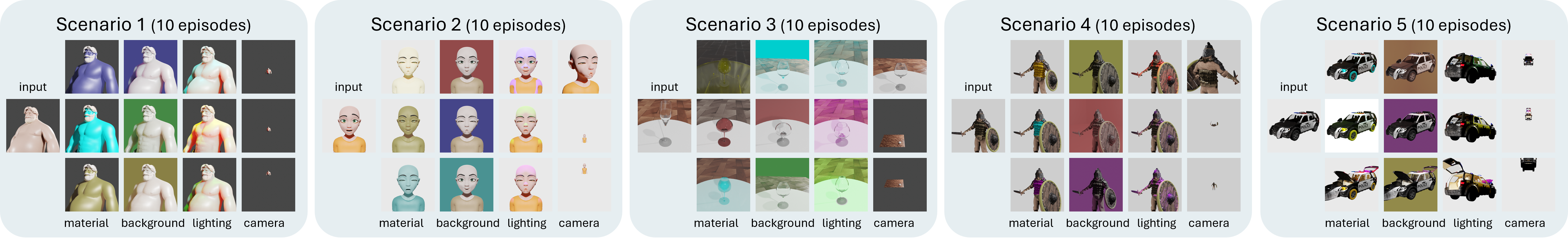}
    \caption{Multi-Tasking Benchmark. Five scenarios (\textbf{50} episodes in total) with editable objects are selected, each paired with a randomly generated prompt to evaluate agent performance on multi-task editing within a single trial. Each trial involves coordinated edits of shape keys, materials, background, lighting, and camera settings according to the prompt.}
    \label{fig:multitask}
    \vspace{-0.25cm}
\end{figure*}

\subsection{Prompt Editing Test}

Prior evaluations of 3D editing agents such as BlenderAlchemy have primarily emphasized precision under controlled start--goal configurations\cite{huang2024blenderalchemy}. While such benchmarks effectively probe an LLM agent’s perception and code organization capabilities within closed environments, they omit the broader spectrum of user experience in realistic, open-ended editing scenarios \cite{gu2025blendergym}. 
To further assess scene understanding and to exploit the multimodal reasoning strengths of vision-language models, we conduct two complementary evaluation protocols: text-prompt editing and visual-prompt editing.  

\paragraph{Text Prompt Editing.}
In the text-prompt editing setting, we follow the design paradigm established by BlenderAlchemy: the user provides a concise natural-language instruction, the agent executes a single edit, and the semantic consistency between the prompt and the rendered output is then quantified. 
Specifically, we benchmarked \textbf{85} representative scenes from BlenderGym, each containing a distinct shape-key object. For each scene, 5 sets of textual prompts were prepared, and 15 trials were conducted per prompt. The rendered outputs are then assessed along 5 evaluation dimensions (shape keys, materials, lighting, background, and camera settings) using CLIP-based alignment scores, which quantify the correspondence between the textual instructions and the generated images.

For each instance $i$, let $I_i$ be the rendered image and $t_i$ the target instruction. 
Let $v_i=\phi(I_i)$ and $u_i=\psi(t_i)$ be the unit–$\ell_2$ CLIP embeddings of $I_i$ and $t_i$, where $\phi$ and $\psi$ are the fixed image/text encoders. 
Specifically, we define the CLIP score as the cosine similarity:
\begin{align}
S_i \;=\; \frac{\langle v_i,\,u_i\rangle}{\|v_i\|_2\,\|u_i\|_2} \;=\; v_i^\top u_i. 
\label{eq:clip_textscore}
\end{align}
Note that both embeddings in Eq. \ref{eq:clip_textscore} are unnormalized to ensure $S_i$ reduces to the inner product (a real-valued scalar, higher is better), following BlenderAlchemy\cite{huang2024blenderalchemy}. 

As shown in Table~\ref{tab:prompt}, EZBlender achieves CLIP scores comparable to BlenderAlchemy and BlenderGPT, and even outperforms them on the shape-key editing task. This indicates that our agent design preserves overall editing quality without noticeable degradation. We note that although BlenderAlchemy is highly effective for task-specific editing, it cannot handle background editing or camera posing due to the lack of predefined agent profiles for these tasks.

\begin{table}[htbp]
    \centering
    \resizebox{1\linewidth}{!}{
    \begin{tabular}{>{\raggedright\arraybackslash}p{0.15\linewidth}ccccc}
        \toprule
        \multicolumn{6}{c}{\textbf{Text-Prompt Editing (CLIP Score $\uparrow$)}} \\
        \midrule
        Method & Shapekey & Material & Background & Lighting & Camera \\
        \midrule
        BlenderGPT& 26.67 & 24.94 & 23.86 & 20.41 & \cellcolor{lime!70}28.33\\
        BlenderAlch.& 28.52 & \cellcolor{lime!70}29.92&   --  & \cellcolor{lime!70}27.44&   --  \\
        EZBlender       & \cellcolor{lime!70}30.21& \cellcolor{lime!30}29.48& \cellcolor{lime!70}28.30& \cellcolor{lime!30}24.06& \cellcolor{lime!30}27.86\\
        \midrule
        \multicolumn{6}{c}{\textbf{Visual-Prompt Editing (CLIP Sim. $\uparrow$)}} \\
        \midrule
        Method & Blend Shape & Placement & Geometry & Lighting & Material \\
        \midrule
        BlenderGPT      & 0.9768 & 0.9482 & 0.9855 & 0.9731 & 0.9806 \\
        BlenderAlch.& 0.9795 & \cellcolor{lime!70}0.9696& \cellcolor{lime!70}0.9914& \cellcolor{lime!70}0.9976& \cellcolor{lime!70}0.9910\\
        EZBlender       & \cellcolor{lime!70}0.9816& \cellcolor{lime!30}0.9559& \cellcolor{lime!30}0.9895& \cellcolor{lime!30}0.9859& \cellcolor{lime!30}0.9876\\
        \bottomrule
    \end{tabular}}
    \caption{\textbf{Prompt Editing Test.} Comparison of text-prompt editing and visual-prompt editing (deeper color is better). 
    }
    \label{tab:prompt}
\end{table}

\paragraph{Visual Prompt Editing.}
The visual prompt editing setting is designed to better capture user intent in complex editing scenarios, as textual descriptions alone often lack precision and omit aesthetic or stylistic nuance \cite{he2024freestyle}. 
Furthermore, visual exemplars (e.g., annotated screenshots or reference images) usually convey explicit constraints on composition, geometry, and illumination. 

For the visual-prompt editing evaluation, we adopt the BlenderGym VLM benchmark \cite{gu2025blendergym}, which reflects actual performance in an open-platform challenge. In this setting, a reference image is provided, and the agent is required to modify the working scene so that the rendered output matches the reference image as closely as possible. This task directly evaluates the precision and reliability of the agent in reproducing fine-grained visual details.

In this setting, the agent is guided by a target (ground-truth) image \(I_i\) and produces a rendered output \(\hat I_i\).
Let \(p_i=\mathrm{norm}\!\big(\phi(I_i)\big)\) and \(v_i=\mathrm{norm}\!\big(\phi(\hat I_i)\big)\) be their unit–\(\ell_2\) CLIP\cite{radford2021learning} embeddings, where \(\phi\) is the fixed image encoder.
We define the visual CLIP similarity as:
\begin{equation}
s_i^{\mathrm{vis}} \;=\; \frac{\langle v_i,\,p_i\rangle}{\|v_i\|_2\,\|p_i\|_2} \;=\; v_i^\top p_i \;\in[-1,1].
\label{eq:clipsim_visual}
\end{equation}
Note that both embeddings in Eq. \ref{eq:clipsim_visual} are normalized to ensure $s_i$ is a cosine-similarity scalar bounded in $[-1,1]$. In particular, although $s_i=1$ does not imply pixel-level identity, it reflects maximal proximity in the CLIP embedding space and thus serves as a faithful proxy for perceptual–aesthetic similarity between the two images \cite{gu2025blendergym}.

Specifically, in the visual-prompt editing evaluation, EZBlender achieves performance comparable to the baselines. We attribute this outcome to improved multi-task coordination and the use of edge autonomy, which allows sub-agents to flexibly refine local components and thereby align the rendered outputs more faithfully with the user’s intent, as shown in Table \ref{tab:prompt}.

\subsection{Multi-Tasking Benchmark}

\begin{figure}[htbp]
    \centering
    \includegraphics[width=1\linewidth]{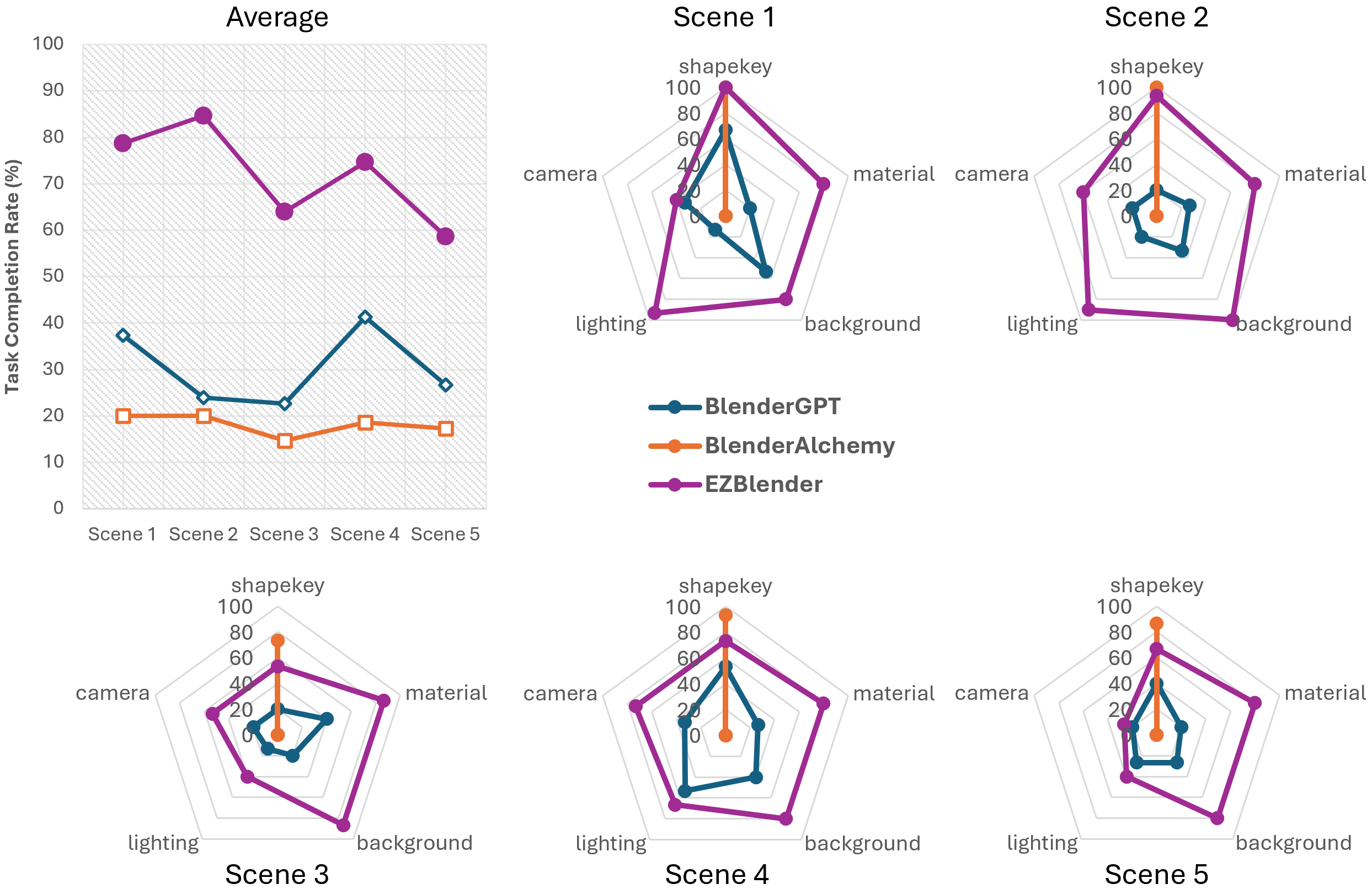}
    \caption{Multi-Tasking benchmark results. Each map shows the task completion rate of different methods for a given scene, while the average plot summarizes overall completion rates across all scenes.}
    \label{fig:multi_plot}
\end{figure}

While the prompt editing experiment focuses on editing precision, many real-world 3D editing scenarios require compositional updates across shape keys, materials, lighting, background, and camera settings. To evaluate this capability, we introduce a Multi-Tasking Benchmark, which measures the task completion rate under natural language prompts that bundle heterogeneous edits.

We design 5 representative scenarios (including \textbf{50} episodes), each built on a predefined editable base mesh (\emph{e.g.}, a human body or a toy figure), as shown in Figure \ref{fig:multitask}. 
For each scene, we conduct 15 trials. In every trial, a random natural language prompt is generated, where each prompt simultaneously specifies 5 concurrent sub-tasks drawn from different domains (shape keys, materials, lighting, and background). 
This setup ensures that editing performance is evaluated in a task-wise and compositional manner.

Specifically, the overall task completion rate (TCR) is defined as:
\begin{align}
    \mathrm{TCR} \;=\; \frac{1}{N}\sum_{i=1}^{N} n_i,
\end{align}
where $N$ represents the total number of tasks, and $n_i$ represents the $i_{th}$ sub-task.
Since each prompt specifies $N$ sub-tasks (e.g., shape key edit, color swap, lighting style change). The agent applies the edits and produces one diagnostic render per sub-task. 
For verification, we use a closed-set CLIP check: compare the rendered image $I_i$ against a small list $\mathcal{V}_i$ of text candidates $t_i$
(e.g., red lighting, blue lighting, green lighting) and take the top-1:
\begin{align}
    \hat{t_i}=\arg\max_{t_i\in\mathcal{V}_i}\ \mathrm{CLIP}\!\big(I_i,\ t_i\big),
\ n_i = 1 \text{ if }\ \hat{t_i}=t_i,
\end{align}
where $\hat{t_i}$ represents the result of CLIP classification of the sub-task $n_i$.
If the predicted result $\hat{t_i}$ is equal to the ground truth label $t_i$, that sub-task is counted as correct; otherwise, it is a miss. Notably, renders that fail also count as misses.

As shown in Figure \ref{fig:multi_plot}, BlenderGPT, despite its simple design, demonstrates strong multi-task capability across most scenes. 
However, its performance on lighting and background is notably weaker. Our observations indicate that the emitted lighting-adjustment code and background-editing code often conflict and are produced without task-specific constraints, leading to failed renders or invalid executions.

In contrast, BlenderAlchemy, while stable and occasionally excelling in specific tasks, fails to generalize to simultaneous multi-editing due to its design trade-offs against generalizability. 
Our proposed EZBlender achieves comparable capability while attaining a higher task completion rate, validating the effectiveness of our edge-autonomy design, in which each sub-agent is specialized for a distinct editing domain, as shown in Table \ref{tab:multitask}. 
It is noteworthy that among all scenes, camera posing and lighting remain the most challenging tasks, largely due to current VLMs' lack of precise 3D spatial perception and closed-loop photometric feedback from the renderer.

\begin{table}[htbp]
    \centering
    \resizebox{0.9\linewidth}{!}{
    \begin{tabular}{lccccc}
        \toprule
        \textbf{TCR (\%) $\uparrow$}& \textbf{S1} & \textbf{S2} & \textbf{S3} & \textbf{S4} & \textbf{S5} \\
        \midrule
        BlenderGPT      & 37.33 & 23.99 & 22.67 & 41.33 & 26.66 \\
        BlenderAlchemy  & 20.00 & 20.00 & 14.67 & 18.67 & 17.33 \\
        EZBlender       & \cellcolor{lime!70}78.67& \cellcolor{lime!70}84.67& \cellcolor{lime!70}60.66& \cellcolor{lime!70}61.33& \cellcolor{lime!70}58.66\\
        \bottomrule
    \end{tabular}}
    \caption{\textbf{Task Completion Rate (TCR) Benchmark.} Completion rates of different Blender-based editing systems across five scenarios with \textbf{50} episodes (deeper color is better). 
    }
    \vspace{-0.5cm}
    \label{tab:multitask}
\end{table}


\subsection{Latency Test}

\begin{figure}[htbp]
    \centering
    \includegraphics[width=1\linewidth]{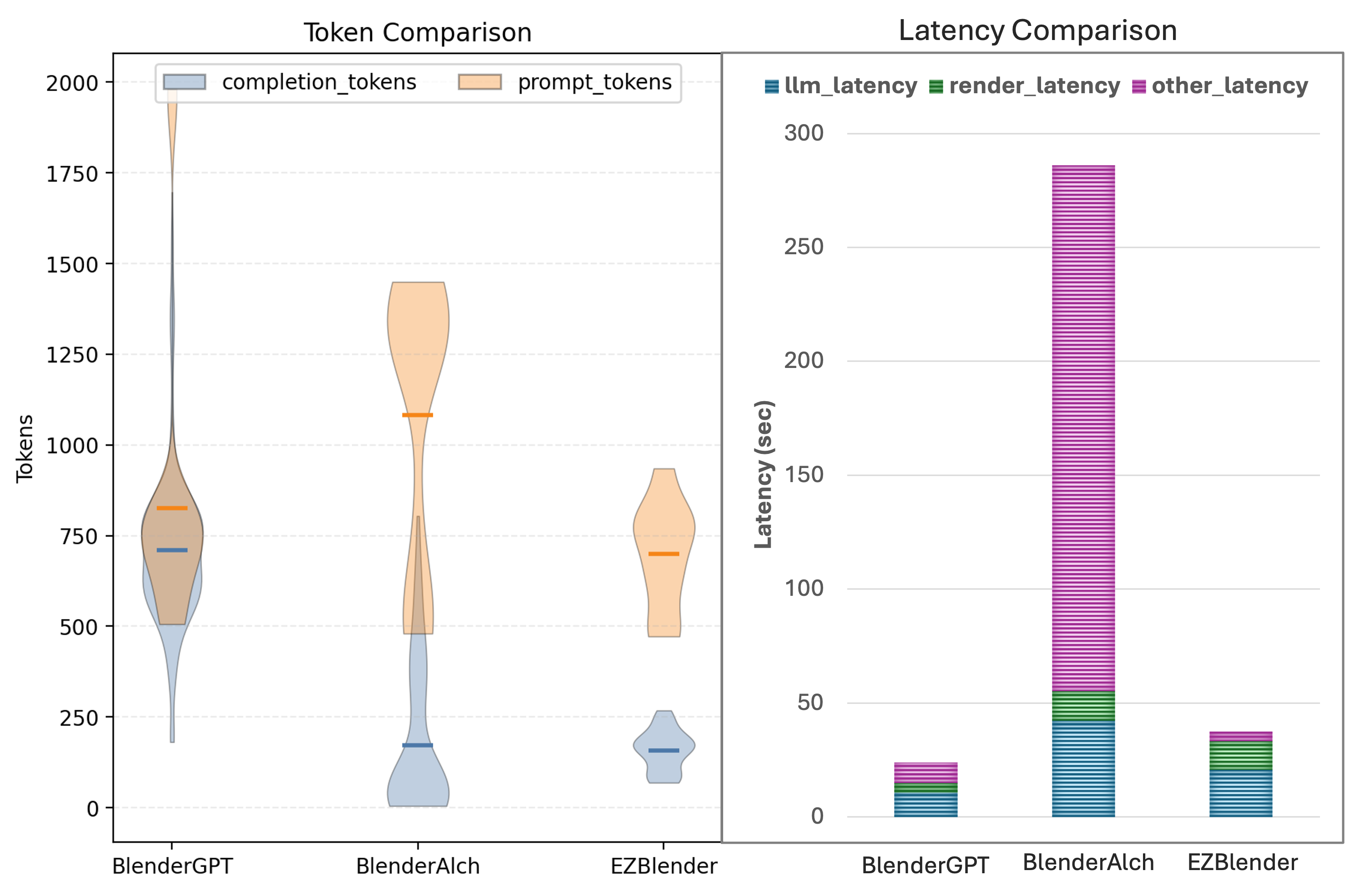}
    \caption{\textbf{Comparison of token consumption and system latency.} (Left) Distribution of prompt and completion tokens across different methods. (Right) Breakdown of system latency into LLM, rendering, and other overhead. 
    }
    \label{fig:latency}
    \vspace{-0.5cm}
\end{figure}

In addition to task precision and completion, responsiveness is also a critical factor for practical 3D editing agents \cite{kim2025cost}. 
To evaluate this aspect, we conduct a latency study of our proposed EZBlender with existing methods. For a fair assessment, we measure the latency of a one-time call (single-time editing request). Specifically, each method is evaluated on 15 distinct single-task prompts, and the average latency is reported. Furthermore, both BlenderAlchemy and the proposed EZBlender are limited to 5 iterations of refinement. 

Latency is further decomposed into three components:
\begin{itemize}
    \item \textbf{Agent latency}: time consumed by LLM reasoning and code generation. For iterative systems such as BlenderAlchemy, this includes repeated planning and refinement steps.
    \item \textbf{Rendering latency}: time spent on Blender execution and image rendering, measured for both low-resolution previews and final outputs. 
    \item \textbf{Other overhead}: includes patch merging, schema validation, self-checks, and I/O operations.
\end{itemize}

\begin{table}[htbp]
    \centering
    \resizebox{0.9\linewidth}{!}{
    \begin{tabular}{lcccc}
        \toprule
        \textbf{Method / Latency (sec) $\downarrow$}& \textbf{LLM} & \textbf{Render} & \textbf{Other} & \textbf{Total} \\
        \midrule
        BlenderGPT      & \cellcolor{lime!70}10.63&  \cellcolor{lime!70}4.16&   9.06 &  \cellcolor{lime!70}23.85\\
        BlenderAlchemy  & 42.29 & 12.88 & 230.91 & 286.08 \\
        Ours            & \cellcolor{lime!30}20.58& \cellcolor{lime!30}12.66&   \cellcolor{lime!70}4.11&  \cellcolor{lime!30}37.35\\
        \bottomrule
    \end{tabular}}
    \caption{\textbf{Latency comparison across different methods.} Latency is split into LLM, rendering, and other overhead (deeper color is better; unit in seconds).
    }
    \label{tab:latency}
\end{table}

As shown in Figure \ref{fig:latency}, BlenderAlchemy exhibits substantial latency issues. Interestingly, its LLM response time and rendering time are comparable to those of the other methods, while the overall speed is significantly hindered by additional system overhead 
due to its tune-leap design \cite{huang2024blenderalchemy}.
In contrast, BlenderGPT achieves the lowest latency, typically responding to user requests and completing a full rendering within a few seconds. Notably, our proposed EZBlender achieves comparable speed to BlenderGPT, with only a slight increase in LLM processing time. The additional cost stems primarily from reasoning at the LLM side, while the agent’s internal decision-making introduces minimal overhead, as reflected in Table \ref{tab:latency}. 
As a result, our parallel design can significantly improve system responsiveness and preserve the overall user experience.

\subsection{Discussion}

\paragraph{Ablation studies.}
To validate the proposed design, we conduct ablation studies to isolate the contributions of each implementation and test the EZBlender under such conditions.  

\begin{table}[htbp]
    \centering
    \resizebox{1\linewidth}{!}{
    \begin{tabular}{lcccc}
        \toprule
        \textbf{Method} 
        & \makecell{\textbf{Text Prompt} \\ \textbf{(CLIP Score$\uparrow$)}}& \makecell{\textbf{Visual Prompt} \\ \textbf{(CLIP Sim.$\uparrow$)}}& \makecell{\textbf{Multi-Tasking} \\ \textbf{(\%) $\uparrow$}}& \makecell{\textbf{Latency} \\ \textbf{(ms) $\downarrow$}}\\
        \midrule
        BlenderGPT              & 24.37& 0.9728 & 30.40 & 23.85 \\
        BlenderAlchemy          & 28.62& 0.9858 & 18.13 & 286.08 \\\midrule
        EZBlender (GPT4o)& 27.61& 0.9801 & 72.13 & 37.35 \\
        EZB. (GPT4o-mini)& 26.80 & 0.9755 & 51.60 & 39.61 \\
 EZB. (GPT4.1-nano)& 24.89 & 0.9699 & 45.66 &48.46 \\\midrule
        EZB. (w/o Reasoning)& 24.45 & 0.9702 & 65.49 & 35.47 \\
        EZB. (w/o Autonomy)& 23.37 & 0.9696 & 33.43 & 30.07 \\
        \bottomrule
    \end{tabular}}
    \caption{Ablation Study. We use EZB. to represent the proposed EZBlender method.}
    \label{tab:ablation}
    \vspace{-0.25cm}
\end{table}

First, we replace the default LLM with lighter variants (e.g., GPT-4o-mini, GPT-4.1-nano) under the same protocol. 
Our observations show a consistent downward shift in prompts that require long-range scene understanding (multi-constraint lighting, coupled camera–material edits). 
Interestingly, even the most lightweight model achieves a higher completion rate than BlenderGPT, as presented in Table \ref{tab:ablation}.
This indicates that our framework benefits from stronger language grounding but remains functional across backbones.

We further ablate the Planning Agent by retaining only the specialized sub-agents with execution autonomy. Consistent with our earlier observations, removing task-aware reasoning may degrade performance on complex prompts and cluttered scenes, where 
the errors usually concentrate on the separation of lighting and background, and the consistency between the modules (Table \ref{tab:ablation}). This degradation underscores the role of task-aware reasoning in regularizing ambiguous instructions into an executable specification.


Lastly, we forbid local self-refinement at the sub-agent level while keeping the planner intact. The system regresses toward BlenderGPT-like behavior, producing valid but brittle one-shot edits with a lower task completion rate and weaker text/visual alignment, as shown in Table \ref{tab:ablation}. This indicates that the proposed edge autonomy is critical for recovering from minor inconsistencies without costly planner round-trips.

\textbf{Token Consumption}
Although our primary focus is on responsiveness and task completion, token efficiency is also critical for practical deployment. 
We therefore analyze token usage across different stages of the pipeline. 

As shown in Figure \ref{fig:latency}, BlenderGPT exhibits the highest variance in token usage (200–1700 tokens per edit), with completion tokens occasionally exceeding 1700. We attribute this instability to the absence of a fixed agent template and consistent reasoning pattern, which induces large fluctuations across nominally similar tasks. 
In contrast, BlenderAlchemy performs more stably but is substantially more token-intensive: its frequent edit–refine cycles and multi-turn reasoning lead to broad oscillations (0–1400 tokens per edit), with prompt contexts often exceeding 1000 tokens, as shown in Table \ref{tab:token}. 
Meanwhile, EZBlender sustains a narrower and lower token profile (typically prompt \textless 1000, completion \textless 300) due to compact plan specifications and limited, localized refinements at execution. While its best-case single-edit token cost can be slightly higher than BlenderGPT, the long-horizon consumption is markedly more stable and economically predictable. 
\begin{table}[htbp]
    \centering
    \renewcommand{\arraystretch}{1.3}
    \resizebox{1\linewidth}{!}{
    \begin{tabular}{lcccc}
        \toprule
        \textbf{Method} & \textbf{Prompt $\downarrow$} & \textbf{Completion $\downarrow$}& \textbf{Total $\downarrow$}& \textbf{Est. Costs (\$) $\downarrow$}\\
        \midrule
        BlenderGPT              &   \cellcolor{lime!70}768&   \cellcolor{lime!70}931&  \cellcolor{lime!70}1,699& 0.0112 \\
        BlenderAlchemy          & 15,262 &  2,543 & 17,805  & 0.0636 \\
        EZBlender (GPT4o)&  \cellcolor{lime!30}4,618&  \cellcolor{lime!30}1,251&  \cellcolor{lime!30}5,869& 0.0241 \\
 EZBlender (GPT4o-mini)& 4,645 & 1,626 & 6,271  &\cellcolor{lime!30}0.0045\\
        EZBlender (GPT4.1-nano)&  7,694&  2,465&  10,159& \cellcolor{lime!70}0.0018\\
        \bottomrule
    \end{tabular}}
    \caption{Token Consumption Comparison. 
    }
    \vspace{-0.5cm}
    \label{tab:token}
\end{table}

\section{Conclusion}

Editing 3D scenes remains a labor-intensive process, where existing VLM-based agents still suffer from weak multi-tasking support and limited responsiveness to complex user instructions.
By combining planning and reasoning with localized edge autonomy, EZBlender achieves efficient task decomposition while preserving semantic fidelity, thus addressing the critical trade-off between precision and responsiveness in human–AI collaboration. 
Compared with the state-of-the-art 3D editing agents, EZBlender achieves up to a $7$-fold improvement in response speed and reduces token consumption by $67\%$.
Beyond quantitative improvements, our framework also demonstrates strong text–visual prompt alignment capabilities, which is one of the most critical functions required for practical 3D editing.
Furthermore, our user-centered pathway for AI-driven 3D editing offers not only a balanced and predictable workflow but also meaningful feedback for the design of future agentic systems. We believe this work broadens the design space for the 3D community, providing new possibilities that emphasize usability, efficiency, and creative flexibility.

{
    \small
    \bibliographystyle{ieeenat_fullname}
    \bibliography{main}

@String(ICLR = {Int. Conf. Learn. Represent.})

@String(ICLR  = {ICLR})

@article{he2024freestyle,
  title={Freestyle: Free lunch for text-guided style transfer using diffusion models},
  author={He, Feihong and Li, Gang and Zhang, Mengyuan and Yan, Leilei and Si, Lingyu and Li, Fanzhang},
  journal={arXiv preprint arXiv:2401.15636},
  year={2024}
}

@inproceedings{yao2023react,
  title={React: Synergizing reasoning and acting in language models},
  author={Yao, Shunyu and Zhao, Jeffrey and Yu, Dian and Du, Nan and Shafran, Izhak and Narasimhan, Karthik and Cao, Yuan},
  booktitle={International Conference on Learning Representations (ICLR)},
  year={2023}
}

@article{liu2024deepseek,
  title={Deepseek-v3 technical report},
  author={Liu, Aixin and Feng, Bei and Xue, Bing and Wang, Bingxuan and Wu, Bochao and Lu, Chengda and Zhao, Chenggang and Deng, Chengqi and Zhang, Chenyu and Ruan, Chong and others},
  journal={arXiv preprint arXiv:2412.19437},
  year={2024}
}

@article{PlanAndSolve,
  title={Plan-and-solve prompting: Improving zero-shot chain-of-thought reasoning by large language models},
  author={Wang, Lei and Xu, Wanyu and Lan, Yihuai and Hu, Zhiqiang and Lan, Yunshi and Lee, Roy Ka-Wei and Lim, Ee-Peng},
  journal={arXiv preprint arXiv:2305.04091},
  year={2023}
}

@article{shinn2023reflexion,
  title={Reflexion: Language agents with verbal reinforcement learning},
  author={Shinn, Noah and Cassano, Federico and Gopinath, Ashwin and Narasimhan, Karthik and Yao, Shunyu},
  journal={Advances in Neural Information Processing Systems},
  volume={36},
  pages={8634--8652},
  year={2023}
}

@inproceedings{raistrick2023infinite,
  title={Infinite photorealistic worlds using procedural generation},
  author={Raistrick, Alexander and Lipson, Lahav and Ma, Zeyu and Mei, Lingjie and Wang, Mingzhe and Zuo, Yiming and Kayan, Karhan and Wen, Hongyu and Han, Beining and Wang, Yihan and others},
  booktitle={Proceedings of the IEEE/CVF conference on computer vision and pattern recognition},
  pages={12630--12641},
  year={2023}
}

@article{wang2023voyager,
  title={Voyager: An open-ended embodied agent with large language models},
  author={Wang, Guanzhi and Xie, Yuqi and Jiang, Yunfan and Mandlekar, Ajay and Xiao, Chaowei and Zhu, Yuke and Fan, Linxi and Anandkumar, Anima},
  journal={arXiv preprint arXiv:2305.16291},
  year={2023}
}

@article{schmidgall2025agent,
  title={Agent laboratory: Using llm agents as research assistants},
  author={Schmidgall, Samuel and Su, Yusheng and Wang, Ze and Sun, Ximeng and Wu, Jialian and Yu, Xiaodong and Liu, Jiang and Moor, Michael and Liu, Zicheng and Barsoum, Emad},
  journal={arXiv preprint arXiv:2501.04227},
  year={2025}
}

@article{yang2024scenecraft,
  title={SceneCraft: Layout-guided 3D scene generation},
  author={Yang, Xiuyu and Man, Yunze and Chen, Junkun and Wang, Yu-Xiong},
  journal={Advances in Neural Information Processing Systems},
  volume={37},
  pages={82060--82084},
  year={2024}
}

@article{lu2025ll3m,
  title={LL3M: Large Language 3D Modelers},
  author={Lu, Sining and Chen, Guan and Dinh, Nam Anh and Lang, Itai and Holtzman, Ari and Hanocka, Rana},
  journal={arXiv preprint arXiv:2508.08228},
  year={2025}
}

@inproceedings{huang2024blenderalchemy,
  title={Blenderalchemy: Editing 3d graphics with vision-language models},
  author={Huang, Ian and Yang, Guandao and Guibas, Leonidas},
  booktitle={European Conference on Computer Vision},
  pages={297--314},
  year={2024},
  organization={Springer}
}

@inproceedings{yi2024gaussiandreamer,
  title={Gaussiandreamer: Fast generation from text to 3d gaussians by bridging 2d and 3d diffusion models},
  author={Yi, Taoran and Fang, Jiemin and Wang, Junjie and Wu, Guanjun and Xie, Lingxi and Zhang, Xiaopeng and Liu, Wenyu and Tian, Qi and Wang, Xinggang},
  booktitle={Proceedings of the IEEE/CVF Conference on Computer Vision and Pattern Recognition},
  pages={6796--6807},
  year={2024}
}

@inproceedings{gu2025blendergym,
  title={BlenderGym: Benchmarking Foundational Model Systems for Graphics Editing},
  author={Gu, Yunqi and Huang, Ian and Je, Jihyeon and Yang, Guandao and Guibas, Leonidas},
  booktitle={Proceedings of the Computer Vision and Pattern Recognition Conference},
  pages={18574--18583},
  year={2025}
}

@article{kim2025cost,
  title={The Cost of Dynamic Reasoning: Demystifying AI Agents and Test-Time Scaling from an AI Infrastructure Perspective},
  author={Kim, Jiin and Shin, Byeongjun and Chung, Jinha and Rhu, Minsoo},
  journal={arXiv preprint arXiv:2506.04301},
  year={2025}
}

@article{dong2025talk,
  title={Talk Before You Retrieve: Agent-Led Discussions for Better RAG in Medical QA},
  author={Dong, Xuanzhao and Zhu, Wenhui and Wang, Hao and Chen, Xiwen and Qiu, Peijie and Yin, Rui and Su, Yi and Wang, Yalin},
  journal={arXiv preprint arXiv:2504.21252},
  year={2025}
}

@misc{BlenderGPT,
  author       = {gd3kr},
  title        = {BlenderGPT},
  year         = {2023},
  publisher    = {GitHub},
  journal      = {GitHub repository},
  howpublished = {\url{https://github.com/gd3kr/BlenderGPT}},
}

@article{chen2025symbolic,
  title={Symbolic Graphics Programming with Large Language Models},
  author={Chen, Yamei and Zhang, Haoquan and Huang, Yangyi and Qiu, Zeju and Zhang, Kaipeng and Wen, Yandong and Liu, Weiyang},
  journal={arXiv preprint arXiv:2509.05208},
  year={2025}
}

@misc{superdesign,
  author       = {superdesigndev},
  title        = {superdesign},
  year         = {2023},
  publisher    = {GitHub},
  journal      = {GitHub repository},
  howpublished = {\url{https://github.com/superdesigndev/superdesign}},
}

@inproceedings{lv2024gpt4motion,
  title={Gpt4motion: Scripting physical motions in text-to-video generation via blender-oriented gpt planning},
  author={Lv, Jiaxi and Huang, Yi and Yan, Mingfu and Huang, Jiancheng and Liu, Jianzhuang and Liu, Yifan and Wen, Yafei and Chen, Xiaoxin and Chen, Shifeng},
  booktitle={Proceedings of the IEEE/CVF conference on computer vision and pattern recognition},
  pages={1430--1440},
  year={2024}
}

@misc{Tripo3D,
  author       = {Tripo3D},
  title        = {3D Model Generator},
  year         = {2023},
  howpublished = {\url{https://www.tripo3d.ai/blog/3d-model-generator}},
  note         = {Accessed: 2025-09-20}
}

@misc{langgraphSupervisor,
  author       = {langchain-ai},
  title        = {langgraph-supervisor-py},
  year         = {2024},
  publisher    = {GitHub},
  journal      = {GitHub repository},
  howpublished = {\url{https://github.com/langchain-ai/langgraph-supervisor-py}},
  note         = {Accessed: 2025-09-20}
}

@inproceedings{wu2024autogen,
  title={Autogen: Enabling next-gen LLM applications via multi-agent conversations},
  author={Wu, Qingyun and Bansal, Gagan and Zhang, Jieyu and Wu, Yiran and Li, Beibin and Zhu, Erkang and Jiang, Li and Zhang, Xiaoyun and Zhang, Shaokun and Liu, Jiale and others},
  booktitle={First Conference on Language Modeling},
  year={2024}
}

@misc{AutoGPT,
  author       = {Significant Gravitas},
  title        = {AutoGPT},
  year         = {2023},
  publisher    = {GitHub},
  journal      = {GitHub repository},
  howpublished = {\url{https://github.com/Significant-Gravitas/AutoGPT}},
}

@misc{Manus,
  author       = {Manus},
  title        = {Manus},
  year         = {2023},
  howpublished = {\url{https://manus.im/}},
  note         = {Accessed: 2025-09-20}
}

@inproceedings{li2023camel,
  title={CAMEL: Communicative Agents for "Mind" Exploration of Large Language Model Society},
  author={Li, Guohao and Hammoud, Hasan Abed Al Kader and Itani, Hani and Khizbullin, Dmitrii and Ghanem, Bernard},
  booktitle={Thirty-seventh Conference on Neural Information Processing Systems},
  year={2023}
}

@misc{OpenAIChatGPTAgent,
  author       = {OpenAI},
  title        = {ChatGPT Agent},
  year         = {2024},
  howpublished = {\url{https://help.openai.com/en/articles/11752874-chatgpt-agent}},
  note         = {Accessed: 2025-09-20}
}

@misc{NanoBananaAI,
  author       = {NanoBanana},
  title        = {NanoBanana AI},
  year         = {2025},
  howpublished = {\url{https://nanobanana.ai/}},
  note         = {Accessed: 2025-09-20}
}

@article{michel2023object,
  title={Object 3dit: Language-guided 3d-aware image editing},
  author={Michel, Oscar and Bhattad, Anand and VanderBilt, Eli and Krishna, Ranjay and Kembhavi, Aniruddha and Gupta, Tanmay},
  journal={Advances in Neural Information Processing Systems},
  volume={36},
  pages={3497--3516},
  year={2023}
}

@inproceedings{sun20253d,
  title={3d-gpt: Procedural 3d modeling with large language models},
  author={Sun, Chunyi and Han, Junlin and Deng, Weijian and Wang, Xinlong and Qin, Zishan and Gould, Stephen},
  booktitle={2025 International Conference on 3D Vision (3DV)},
  pages={1253--1263},
  year={2025},
  organization={IEEE}
}

@article{nie2025ermv,
  title={ERMV: Editing 4D Robotic Multi-view images to enhance embodied agents},
  author={Nie, Chang and Wang, Guangming and Lie, Zhe and Wang, Hesheng},
  journal={arXiv preprint arXiv:2507.17462},
  year={2025}
}

@article{liu2025worldcraft,
  title={WorldCraft: Photo-realistic 3D world creation and customization via LLM agents},
  author={Liu, Xinhang and Tang, Chi-Keung and Tai, Yu-Wing},
  journal={arXiv preprint arXiv:2502.15601},
  year={2025}
}

@book{hamdani20233d,
  title={3D Environment Design with Blender: Enhance your modeling, texturing, and lighting skills to create realistic 3D scenes},
  author={Hamdani, Abdelilah},
  year={2023},
  publisher={Packt Publishing Ltd}
}

@inproceedings{zhuang2023dreameditor,
  title={Dreameditor: Text-driven 3d scene editing with neural fields},
  author={Zhuang, Jingyu and Wang, Chen and Lin, Liang and Liu, Lingjie and Li, Guanbin},
  booktitle={SIGGRAPH Asia 2023 Conference Papers},
  pages={1--10},
  year={2023}
}

@inproceedings{karim2024free,
  title={Free-editor: zero-shot text-driven 3D scene editing},
  author={Karim, Nazmul and Iqbal, Hasan and Khalid, Umar and Chen, Chen and Hua, Jing},
  booktitle={European Conference on Computer Vision},
  pages={436--453},
  year={2024},
  organization={Springer}
}

@inproceedings{cai20233description,
  title={3Description: An Intuitive Human-AI Collaborative 3D Modeling Approach},
  author={Cai, Zhuodi},
  booktitle={Proceedings of the 11th International Conference on Digital and Interactive Arts},
  pages={1--5},
  year={2023}
}

@article{jun2023shap,
  title={Shap-e: Generating conditional 3d implicit functions},
  author={Jun, Heewoo and Nichol, Alex},
  journal={arXiv preprint arXiv:2305.02463},
  year={2023}
}

@inproceedings{lin2023magic3d,
  title={Magic3d: High-resolution text-to-3d content creation},
  author={Lin, Chen-Hsuan and Gao, Jun and Tang, Luming and Takikawa, Towaki and Zeng, Xiaohui and Huang, Xun and Kreis, Karsten and Fidler, Sanja and Liu, Ming-Yu and Lin, Tsung-Yi},
  booktitle={Proceedings of the IEEE/CVF conference on computer vision and pattern recognition},
  pages={300--309},
  year={2023}
}

@article{gao2022get3d,
  title={Get3d: A generative model of high quality 3d textured shapes learned from images},
  author={Gao, Jun and Shen, Tianchang and Wang, Zian and Chen, Wenzheng and Yin, Kangxue and Li, Daiqing and Litany, Or and Gojcic, Zan and Fidler, Sanja},
  journal={Advances in neural information processing systems},
  volume={35},
  pages={31841--31854},
  year={2022}
}

@article{li2024advances,
  title={Advances in 3d generation: A survey},
  author={Li, Xiaoyu and Zhang, Qi and Kang, Di and Cheng, Weihao and Gao, Yiming and Zhang, Jingbo and Liang, Zhihao and Liao, Jing and Cao, Yan-Pei and Shan, Ying},
  journal={arXiv preprint arXiv:2401.17807},
  year={2024}
}

@book{flavell2011beginning,
  title={Beginning blender: open source 3d modeling, animation, and game design},
  author={Flavell, Lance},
  year={2011},
  publisher={Apress}
}

@article{hosen2019mastering,
  title={Mastering 3D modeling in blender: from novice to pro},
  author={Hosen, Md Saikat and Ahmmed, Shahed and Dekkati, Sreekanth},
  journal={ABC Research Alert},
  volume={7},
  number={3},
  pages={169--180},
  year={2019}
}

@article{AHMAD20171,
title = {How to launch a successful video game: A framework},
journal = {Entertainment Computing},
volume = {23},
pages = {1-11},
year = {2017},
issn = {1875-9521},
doi = {https://doi.org/10.1016/j.entcom.2017.08.001},
url = {https://www.sciencedirect.com/science/article/pii/S1875952117300861},
author = {Norita B. Ahmad and Salahudin Abdul Rahman Barakji and Tarak Mohamed Abou Shahada and Zeid Ayman Anabtawi},
}

@inproceedings{bhuyan2024gamestreamsr,
  title={GameStreamSR: Enabling neural-augmented game streaming on commodity mobile platforms},
  author={Bhuyan, Sandeepa and Ying, Ziyu and Kandemir, Mahmut T and Gowda, Mahanth and Das, Chita R},
  booktitle={2024 ACM/IEEE 51st Annual International Symposium on Computer Architecture (ISCA)},
  pages={1309--1322},
  year={2024},
  organization={IEEE}
}

@article{banfi2024gaming,
  title={Gaming i, II, and III: Arcades, video game systems, and modern game streaming services},
  author={Banfi, Ryan},
  journal={Games and Culture},
  volume={19},
  number={8},
  pages={1000--1037},
  year={2024},
  publisher={SAGE Publications Sage CA: Los Angeles, CA}
}

@book{alma9976203403606533,
author = {Dunlop, Renee},
address = {New York},
booktitle = {Production pipeline fundamentals for film and game},
isbn = {9780415812290},
language = {eng},
lccn = {2013032579},
publisher = {Focal Press, Taylor Francis Group},
title = {Production pipeline fundamentals for film and game  / edited by Renee Dunlop.},
year = {2014},
}

@article{gollnersony,
  title={SONY 360 REALITY AUDIO/MPEG-H 3D AUDIO},
  author={G{\"o}llner, Louis}
}

@inproceedings{radford2021learning,
  title={Learning transferable visual models from natural language supervision},
  author={Radford, Alec and Kim, Jong Wook and Hallacy, Chris and Ramesh, Aditya and Goh, Gabriel and Agarwal, Sandhini and Sastry, Girish and Askell, Amanda and Mishkin, Pamela and Clark, Jack and others},
  booktitle={International conference on machine learning},
  pages={8748--8763},
  year={2021},
  organization={PMLR}
}

@misc{blender,
  title        = {Blender: Free and Open Source 3D Creation Suite},
  author       = {{Blender Online Community}},
  year         = {2025},
  howpublished = {\url{https://www.blender.org/}},
  note         = {Blender Foundation, Amsterdam}
}

@misc{adk_python,
  title        = {ADK-Python: Agent Development Kit for Building AI Agents},
  author       = {{Google} and Contributors},
  year         = {2025},
  howpublished = {\url{https://github.com/google/adk-python}},
  note         = {commit as of access, Apache-2.0 license}
}

@article{lu2024advances,
  title={Advances in text-guided 3D editing: a survey},
  author={Lu, Lihua and Li, Ruyang and Zhang, Xiaohui and Wei, Hui and Du, Guoguang and Wang, Binqiang},
  journal={Artificial Intelligence Review},
  volume={57},
  number={12},
  pages={321},
  year={2024},
  publisher={Springer}
}

@inproceedings{SODA,
  title={SODA: Spectral Orthogonal Decomposition Adaptation for Diffusion Models},
  author={Zhang, Xinxi and Wen, Song and Han, Ligong and Juefei-Xu, Felix and Srivastava, Akash and Huang, Junzhou and Pavlovic, Vladimir and Wang, Hao and Tao, Molei and Metaxas, Dimitris},
  booktitle={2025 IEEE/CVF Winter Conference on Applications of Computer Vision (WACV)},
  pages={4665--4682},
  year={2025},
  organization={IEEE}
}

@article{FlowStraighter,
  title={Flow Straighter and Faster: Efficient One-Step Generative Modeling via MeanFlow on Rectified Trajectories},
  author={Zhang, Xinxi and Tan, Shiwei and Nguyen, Quang and Dao, Quan and Han, Ligong and He, Xiaoxiao and Zhang, Tunyu and Mrdovic, Alen and Metaxas, Dimitris},
  journal={arXiv preprint arXiv:2511.23342},
  year={2025}
}

@article{BezierSplat,
  title={B$\backslash$'ezier Splatting for Fast and Differentiable Vector Graphics Rendering},
  author={Liu, Xi and Zhou, Chaoyi and Zhao, Nanxuan and Huang, Siyu},
  journal={arXiv preprint arXiv:2503.16424},
  year={2025}
}

@inproceedings{Latent3DGS,
  title={Latent Radiance Fields with 3D-aware 2D Representations},
  author={Zhou, Chaoyi and Liu, Xi and Luo, Feng and Huang, Siyu},
  booktitle={The Thirteenth International Conference on Learning Representations}
}

@article{RetinalGPT,
  title={RetinalGPT: A Retinal Clinical Preference Conversational Assistant Powered by Large Vision-Language Models},
  author={Zhu, Wenhui and Li, Xin and Chen, Xiwen and Qiu, Peijie and Vasa, Vamsi Krishna and Dong, Xuanzhao and Chen, Yanxi and Lepore, Natasha and Dumitrascu, Oana and Su, Yi and others},
  journal={arXiv preprint arXiv:2503.03987},
  year={2025}
}

@article{LLaDAMedV,
  title={Llada-medv: Exploring large language diffusion models for biomedical image understanding},
  author={Dong, Xuanzhao and Zhu, Wenhui and Chen, Xiwen and Wang, Zhipeng and Qiu, Peijie and Tang, Shao and Li, Xin and Wang, Yalin},
  journal={arXiv preprint arXiv:2508.01617},
  year={2025}
}
}

\end{document}